\newcommand{\vecA}{\mbox{\boldmath$A$}}
\newcommand{\vecE}{\mbox{\boldmath$E$}}
\newcommand{\vecB}{\mbox{\boldmath$B$}}

\newcommand{\vece}{\mbox{\boldmath$e$}}

\newcommand{\dfd}{{\rm d}}
\newcommand{\vecr}{\mbox{\boldmath$r$}}
\newcommand{\vecv}{\mbox{\boldmath$v$}}

\newcommand{\vecOmega}{\mbox{\boldmath$\Omega$}}

\documentclass[12pt,a4paper]{article}
\usepackage{graphicx}
\title{Connection between London moment and Meissner effect from
classical electrodynamics}
\author{Hanno Ess\'en\\
Department of Mechanics \\Royal Institute
of Technology  \\ S-100 44 Stockholm, Sweden}
\date{September 14, 2004}
\begin{document}
\maketitle
\begin{abstract}
Theory and experiment on the London moment is reviewed. A simple
mathematical model is motivated and then used to study the
responses of a spherical superconductor to an external field and
to rotation. It reveals a connection between perfect diamagnetism
(Meissner effect) and the London moment. In the model neither of
these are exact but the deviation from $B=0$ internal field in the
former and from $B=(2mc/e)\Omega$ in the latter case is described
by the same dimensionless parameter. Apart from its pedagogical
values the model might throw some light on the controversy
surrounding the correction to the London moment.
\end{abstract}

\section{Introduction}
When a superconductor is rotated with angular velocity $\vecOmega
=(\omega\, {\rm rad/s})\vece_z$ a magnetic field,
\begin{equation}\label{eq.lond.mom}
\vecB = \frac{2mc}{e}\vecOmega,
\end{equation}
with $B= 1.137\cdot 10^{-11}\, \omega
 \;{\rm tesla}$, arises inside it. Here $-e$ is electron charge and $m$
electron mass. This is called the London moment since it was
predicted by Fritz London \cite{london} on the basis of the London
brothers' phenomenological theory of superconductivity, but the
formula was in fact derived much earlier by Becker et al.\
\cite{becker} using the non-viscous electronic liquid model. Since
then various ways of arriving at this formula have been proposed
\cite{rystephanick,hirsch}. The shortest heuristic derivation
postulates that effective forces in the rotating system must
vanish; the field (\ref{eq.lond.mom}) is then needed to cancel the
Coriolis force (Rystephanick \cite{rystephanick2}).

Formula (\ref{eq.lond.mom}) is remarkable since it gives the
electronic charge to mass ratio from macroscopic measurement and
its basic correctness has been experimentally verified by
Hildebrandt \cite{hildebrandt}. It has also been verified that it
is independent of the type of superconductor
\cite{verheijen,sanzari} and of its initial rotational state
\cite{hipkins}. Nowadays it is used in basic physics experiments
\cite{buchman}. This immediately leads to the question of how
accurate it is.

Since replacement of $e$ and $m$ by $Ne$ and $Nm$ leaves formula
(\ref{eq.lond.mom}) invariant it may in fact refer to the charge
to mass ratio of Cooper pairs or of larger groups of electrons
such as the entire superconducting condensate. Based on various
theoretical assumptions one can approach the question of
corrections to (\ref{eq.lond.mom}) and this has been done by
several authors \cite{brady,cabrera,liu,jiang,berger,baym}. The
results do not agree, however;  neither with each other nor with
experiment \cite{tate}. In view of this confusion it may be worth
while to point out that even a very basic classical model of the
phenomenon leads to a correction to London's formula.

We will first motivate heuristically that our model should
qualitatively describe the physics of a superconducting sphere.
After that the model system, and its kinematics, its basic
parameters, and its dynamics, are presented. Only classical
mechanics and electrodynamics is used. Diamagnetism is then
studied within the model and it turns out to be perfect only in
the limit of infinitely many electrons. We finally turn to the
response of the model to rotation and find that the London moment
becomes exact in the same limit that achieved perfect
diamagnetism.

\section{The giant atom idea}
After Meissner's \cite{meissner} discovery in 1933 of the
expulsion of a magnetic field from the superconductor at its phase
transition it was realized that understanding the perfect
diamagnetism might be one clue to a theory of superconductors.
This lead Welker \cite{welker} to the study of superconductors as
giant atoms. He was inspired by Langevin's theory of diamagnetism
for systems of closed shells atoms and ions. In this theory the
external field induces a rotation of the atoms and these rotating
atoms produce a field that opposes the external field. For an
illuminating discussion see Ess\'en \cite{essen89}, see also van
Vleck \cite{vanvleck}. In ordinary metals the magnetic
susceptibility is nearly zero because, as Welker explained, the
diamagnetic effect is exactly balanced by a paramagnetic effect,
the ordering of the electron spins along the external field. In
this way Welker \cite{welker} realized that perfect diamagnetism
requires that there is a gap in the spectrum of the conduction
electrons which is not present in ordinary metals. With this
energy gap the Langevin mechanism can be blown up and the
paramagnetism suppressed. In recent years Hirsch
\cite{hirsch,hirsch1} has advocated the giant atom view of
superconductors, see also Ess\'en \cite{essen00}.

Since the discovery of the Pauli principle is has been realized
that the electrons that participate in conduction of electricity
are the electrons at the surface of the Fermi sea of degenerate
electrons. Electrons inside the surface are not able to change
their state of motion. The relevant electrons are thus those with
the largest energies and velocities \cite{essen95}, essentially
the Fermi energy and Fermi velocity, $v_{\rm F}$. In a normal
metal such electrons are scattered and have a short mean free path
$\Lambda$. The time between collisions are then on average,
$\tau=\Lambda/v_{\rm F}$. As long as the metal is large compared
to $\Lambda$ the conduction electron gas will thus be {\em
homogeneous} throughout the metal.

In a superconductor, on the other hand, Cooper pairs will form,
and at the critical temperature these must be interpreted as
having infinite mean free path, $\Lambda\rightarrow\infty$. When
the mean free path becomes of the same order of magnitude as the
container, the gas can no longer be homogeneous. Instead its
distribution must be strongly influenced by the shape of the
container. In a spherical metal ball of radius $R$ one then gets
an even better analogy with a giant atom. The Cooper pairs can
move freely in the spherical container. Since their electrons must
still must have the largest energy and momenta among the electrons
according to the Pauli exclusion principle this means that they
must spend most of their time near the metal surface. The
centrifugal potential for particles with the Fermi momentum will
be order of magnitude $\sim R^2 p_{\rm F}^2/2mr^2$, and thus most
pairs are pushed to the surface. We will not go more deeply into
this here; we just note that for a superconductor the Fermi
surface and surface of the metal are necessarily close. Already
London \cite{london} states that superconductivity is a {\em
surface phenomenon,} but this nowadays sometimes seems to be
forgotten. The fact that the superconducting condensate is
concentrated near the metal surface is the motivation for the
model presented in the next section.

\section{The model system}
Consider a heavy sphere of radius $R$ with a positive surface
charge $Q$ and surface density $\sigma_+=Q/4\pi R^2$. An
oppositely charged thin spherical shell, of mass $M$, and the same
radius $R$, covers the surface of the sphere but can rotate freely
on it. The system is thus electrically neutral but surface
currents, corresponding to rigid rotation of the negative surface
charge density, $\sigma_-=-\sigma_+$, can flow without dissipation

We now set up the Lagrangian of this system in an external
magnetic field with vector potential $\vecA_{\rm e}$. Since we
safely can neglect radiation in our problem we can use the Darwin
Lagrangian (see Jackson \cite{jackson3}, Ess\'en
\cite{essen96,essen99}), but we skip the relativistic correction
to the kinetic energy as discussed by Ess\'en \cite{essen99}. We
have,
\begin{equation}\label{eq.basic.lagrangian}
L(\vecr_k,\vecv_k) =\frac{1}{2} \sum_{k=1}^N  m_k \vecv_k^2
+\frac{1}{2} \sum_{k=1}^N
 \frac{q_k}{c} \vecv_k\cdot   \vecA_{\rm i}(\vecr_k) +
\sum_{k=1}^N \frac{q_k}{c} \vecv_k\cdot \vecA_{\rm e}(\vecr_k),
\end{equation}
where $\vecA_{\rm i}(\vecr_k)$ is the internal vector potential
from the particles of the system. It is a sum over all particles
except particle number $k$ and the second sum in $L$ is thus a sum
over pair interactions; therefore the factor one half in front.
The important thing in the Darwin formalism is that $\vecA_{\rm
i}$ is divergence free (Coulomb gauge). The last sum is the usual
one representing the interaction with the external vector
potential $\vecA_{\rm e}$.

We will use spherical coordinates ($r,\theta,\varphi$), so the
velocity of a particle fixed on the rotating shell is,
\begin{equation}\label{eq.vel}
\vecv(\theta,\varphi,\dot\varphi) =\dot\varphi\vece_z\times\vecr=
R\sin\theta\,\dot\varphi\,\vece_{\varphi}(\varphi).
\end{equation}
For the kinetic energy we must integrate over the sphere $r=R$,
and we find,
\begin{equation}\label{eq.kin.energy.def}
T=\frac{1}{2}\sum_{k=1}^N  m_k \vecv_k^2 =\frac{1}{2} \int_{S}
\dfd m(\theta,\varphi) \;{\vecv}^2 (\theta,\varphi,\dot\varphi)
=\frac{1}{3}M R^2\dot\varphi^2,
\end{equation}
in agreement with the fact that the moment of inertia of a
spherical shell is $I_z = (2/3)MR^2$.

To find the vector potential of the current from the rotating
shell, with charge $-Q$, is an elementary exercise \cite{good}.
Some useful formulas can be found in Ess\'en \cite{essen89}. At
$r=R$ the result is,
\begin{equation}\label{eq.internal.A}
 \vecA_{\rm i}(\theta,\varphi,\dot\varphi) =
-\frac{\dot\varphi}{c}\frac{Q}{3}\sin\theta\,\vece_{\varphi}(\varphi).
\end{equation}
The self interaction term in the Lagrangian is thus
\begin{equation}\label{eq.Li}
L_{\rm i}=\frac{1}{2c} \int_{S} \dfd q(\theta,\varphi) \; \vecv
(\theta,\varphi,\dot\varphi)\cdot\vecA_{\rm
i}(\theta,\varphi,\dot\varphi) = \frac{RQ^2}{9c^2}\dot\varphi^2,
\end{equation}
and is seen to be similar to the kinetic energy term.

For definiteness we here compute the last term for the case of a
homogeneous external field $\vecB=B_{\rm e}\vece_z$. The vector
potential is then,
\begin{equation}\label{eq.external.A}
 \vecA_{\rm e}(r,\theta,\varphi)
 =\frac{1}{2}B_{\rm e}(-y\vece_x+x\vece_y)=
\frac{1}{2}B_{\rm e} r \sin\theta\,\vece_{\varphi}(\varphi)
\end{equation}
and one thus finds,
\begin{equation}\label{eq.Le}
L_{\rm e} =\frac{1}{c} \int_{S} \dfd q(\theta,\varphi) \; \vecv
(\theta,\varphi,\dot\varphi)\cdot\vecA_{\rm e}(R,\theta,\varphi) =
-\frac{R^2Q}{3c}B_{\rm e} \dot\varphi,
\end{equation}
for the interaction Lagrangian of the rotating spherical shell
with this field. This is the interaction needed to study
diamagnetism. To investigate the London moment below we have to
modify the external field.

Collecting terms we now get get,
\begin{equation}\label{eq.lagr.resultat}
 L(\dot\varphi)= T+L_{\rm i}+L_{\rm e} =\frac{R^2}{3}
 \left[M\left(1+\frac{Q^2}{3RMc^2}
 \right)\dot\varphi^2 - \frac{Q}{c}B_{\rm e} \dot\varphi\right],
\end{equation}
for our Lagrangian. If we use $Q=Ne$, $M=Nm$, and the classical
electron radius $r_e =\frac{e^2}{mc^2}$, we can write,
\begin{equation}\label{eq.def.cal.m}
M\left(1+\frac{Q^2}{3RMc^2}
 \right) = Nm\left(1+\frac{N r_e}{3R}\right)\equiv Nm(1+\epsilon N),
\end{equation}
and rewrite the Lagrangian in the simple form,
\begin{equation}\label{eq.lagr.resultat.simpl}
 L(\dot\varphi)=\frac{NmR^2}{3}
 \left[(1+\epsilon N) \dot\varphi^2 - \frac{e}{mc}B_{\rm e}
 \dot\varphi\right].
\end{equation}
We see that the generalized coordinate $\varphi$ is absent (i.e.\
cyclic) and that the generalized momentum is
\begin{equation}\label{eq.gen.mom}
p_{\varphi}=\frac{\partial L}{\partial \dot\varphi}
=\frac{2NmR^2}{3} \left[(1+\epsilon N) \dot\varphi -
\frac{e}{2mc}B_{\rm e}\right].
\end{equation}
The corresponding Hamiltonian is given by $H=\dot\varphi
p_{\varphi}-L$ and
\begin{equation}\label{eq.hamiltonian}
H(p_{\varphi}) =\frac{3}{4}\frac{N}{m(1+\epsilon
N)}\left(\frac{p_{\varphi}}{NR} +\frac{eR}{3c}B_{\rm e} \right)^2
\end{equation}
is the result of the calculation.

\section{Diamagnetism and Meissner effect}
The Meissner effect \cite{meissner} is strictly speaking the fact
that a superconductor {\em expels} a magnetic field when cooled
below the critical temperature. In this it is different
thermodynamically from a so called perfect conductor which merely
has zero resistance, see Jackson \cite{jackson3}, Pippard
\cite{pippard}. Here we will not discuss thermodynamics and phase
transitions, so we can be a bit sloppy and refer to the Meissner
effect simply as the fact that an external field will not enter
the superconducting body when it is switched on. In short, we will
discuss the {\em perfect diamagnetism} of superconductors.

Let us see what our model system predicts if we take the {\em
initial conditions} to be $\dot\varphi(0)=0$ when the external
field is zero $B_{\rm e}(0)=0$. The equation of motion is, $\dot
p_{\varphi} =
\partial L/\partial \varphi =0$, so the generalized momentum is
conserved. The initial conditions give $p_{\varphi}=0$ and then
Eq. (\ref{eq.gen.mom}) gives,
\begin{equation}\label{eq.solution.dot.phi}
(1+\epsilon N) \dot\varphi(t) =\frac{e}{2mc}B_{\rm e}(t) ,
\end{equation}
at all times. The angular velocity of the shell is completely
determined by the external field at all times. Here this follows
from our conservation law $p_{\varphi}=\,$constant. Becker et al.\
\cite{becker} explains this by saying that the electric field
$\vecE=-(1/c)\,\partial\vecA_{\rm e}/\partial t$ causes
acceleration of the shell .

The rotating shell will of course produce a magnetic field $B_{\rm
i}$ of its own. Inside the shell ($r\le R$) it is homogeneous and
can be read of by comparing Eqs. (\ref{eq.internal.A}) and
(\ref{eq.external.A}). This gives,
\begin{equation}\label{eq.intBi}
B_{\rm i}(t)=-\frac{2}{3}\frac{Q}{R}\frac{\dot\varphi(t)}{c}=
-N\frac{2}{3}\frac{e}{R}\frac{\dot\varphi(t)}{c},
\end{equation}
for the induced field inside the sphere (outside the shell one
finds a pure dipole field \cite{essen98}). Using
(\ref{eq.solution.dot.phi}) this can be expressed in terms of
$B_{\rm e}$. The {\em total field inside the sphere} is then
\begin{equation}\label{eq.Bdia}
B_{\rm dia} = B_{\rm e} + B_{\rm i}=B_{\rm e} \left(
\frac{1}{1+\epsilon N}\right).
\end{equation}
Here $\epsilon$ was defined in (\ref{eq.def.cal.m}) and is
\begin{equation}\label{eq.eps.def}
\epsilon=\frac{r_e}{3R}.
\end{equation}
We see that {\em perfect diamagnetism} ($B_{\rm dia}\rightarrow
0$) corresponds to $N\rightarrow\infty$, so for finite $N$ it can
not be achieved, but it gets better the larger the system.

One notes that our model for diamagnetism here is almost entirely
like the old Langevin theory. The main difference is that we are
not using Larmor's theorem and thus we are {\em not} assuming that
the external field is a {\em weak perturbation,} as is required
for the use of Larmor's  formula \cite{essen00}. Instead
everything is exact within the model. The smallness of ordinary
diamagnetism, when the spheres are atoms, is due to the fact that
$N\sim 10$ and $\epsilon\sim r_e/3a_0\approx 1.78\cdot 10^{-5}$,
where $a_0$ is the Bohr radius. Clearly only a very small
reduction of the external field is possible in this case.

What about the macroscopic superconducting spheres? For $R=1\,$cm
one finds that $\epsilon\approx 10^{-13}$. Does the quantity
$\epsilon N =Nr_e/3R$ grow sufficiently to produce nearly perfect
diamagnetism? One might assume that $N \propto R^3$ but this is
not correct. The conduction electrons and thus also the
superconducting condensate consists of electrons from a thin layer
at the Fermi surface in momentum space. Since this is a
two-dimensional object the number of relevant electrons must obey
$N \propto R^2$ (Ess\'en \cite{essen95}). Incidentally this gives
the physical result that the surface charge density
$\sigma_-=-Ne/4\pi R^2$, of our model, can remain constant as $R$
increases. The simplest possible minimum estimate assumes that
each surface atom contributes one Fermi surface electron and that
only these participate in the condensate. This gives $N\approx
R^2/a_0^2$. We then find that $\epsilon N \approx
(r_e/3R)(R^2/a_0^2)=3.3\cdot 10^5{\rm m}^{-1} R$. For $R=1\,$cm
this gives $\epsilon N \approx 3300$, so macroscopic spheres
should in fact be highly diamagnetic.

\section{Rotation and London Moment}
We now come to the main task of this work. What is the field of a
rotating superconductor? Since our model managed to predict strong
diamagnetism it might also give decent results in this case. The
external field is no longer assumed to be a homogenous field.
Instead we now start rotating the heavy sphere with the positive
surface charge density $\sigma_+=Ne/4\pi R^2$. When this sphere
rotates with angular velocity $\Omega$ it will produce the field,
\begin{equation}\label{eq.B.Omega}
B_{\rm e}(t)= N\frac{2}{3}\frac{e}{R}\frac{\Omega(t)}{c},
\end{equation}
for $r\le R$, in analogy with Eq.\ (\ref{eq.intBi}). (Outside the
sphere it is a dipole field and goes to zero at infinity, just as
the field $B_{\rm i}$ above.)

Assuming initial conditions $\dot\varphi(0)=0$ when $\Omega(0)=0$,
we again get Eq.\ (\ref{eq.solution.dot.phi}) for the induced
angular velocity $\dot\varphi(t)$ of the freely rotating
negatively charged shell. Eq.\ (\ref{eq.solution.dot.phi}) now
relates $\dot\varphi(t)$ and $\Omega(t)$ at all times. To find the
internal (London) field in this case all we have to do is to use
Eq.\ (\ref{eq.Bdia}) and replace $B_{\rm e}$ on the right hand
side with the expression (\ref{eq.B.Omega}). This produces the
result,
\begin{equation}\label{eq.Blond}
B_{\rm Lond} = B_{\rm e} + B_{\rm i}=\frac{2mc}{e}\Omega \left(
\frac{\epsilon N}{1+\epsilon N}\right),
\end{equation}
after some simple algebra. When $N\rightarrow\infty$ this {\em
approaches the London moment} ($B_{\rm
Lond}\rightarrow\frac{2mc}{e}\Omega$) of Eq.\ (\ref{eq.lond.mom}).
Just as was the case above with the perfect diamagnetism we find
that the London moment is exact only in the limit of infinitely
many particles. If we trace the origin of the terms we see that
the extra 1 in the denominator of (\ref{eq.Blond}) is due to the
contribution to inertia from electron mass, while $\epsilon N$
comes from the inductive inertia that reflects the energy cost of
building up a magnetic field. In electric circuit theory one is
used to considering only the inductive inertia. Inertia due to
electron mass is usually negligible in such experiments. In high
precision measurements, however, the electron inertia may play a
role and thus the correction term to the London moment suggested
by Eq.\ (\ref{eq.Blond}) may have to be taken seriously.

\section{Discussion and conclusions}
The beauty of our embarrassingly simple model is that it does not
just give the London moment, as many other oversimplified studies.
Instead it gives the London moment only as a limit for
$N\rightarrow \infty$, {\em and\/} it shows how this limit is
intimately connected with the limit of perfect diamagnetism. This
is no mean achievement for such a small investment and must be
regarded as physics pedagogics at its best.

While most textbooks seem to ignore the London moment there is
still a fair amount of active research in this and related areas
\cite{gawlinski,rojo,fischer}. It has been pointed out that the
universality of the London moment, and its sign in particular,
means that the superconducting charge carriers are always
electrons, not holes \cite{dunne}. If nothing else, this article
would therefore, at least, like to make the theoretical and
experimental fact of the London moment better known. It is just as
remarkable as zero resistivity and perfect diamagnetism, not to
mention  the Josephson effect.


\end{document}